\documentclass[conference]{IEEEtran}
\ifCLASSINFOpdf
\else
\fi

\usepackage{amssymb}
\usepackage{setspace}
\usepackage{graphicx}
\usepackage[tight,footnotesize]{subfigure}
\usepackage{booktabs}
\usepackage{amsthm}
\usepackage{amsmath}
\usepackage{marvosym}
\usepackage{multirow}
\usepackage{caption}
\usepackage{slashbox}
\usepackage{cite}
\usepackage{array}
\def\BibTeX{{\rm B\kern-.05em{\sc i\kern-.025em b}\kern-.08em T\kern-.1667em\lower.7ex\hbox{E}\kern-.125emX}}

\allowdisplaybreaks[4]
\begin{document}
\title{vFAC: Fine-Grained Access Control with Versatility for Cloud Storage}
\author{\IEEEauthorblockN{Jingwei Liu\IEEEauthorrefmark{1},
Huifang Tang\IEEEauthorrefmark{1},
Chaoya Li\IEEEauthorrefmark{1},
Rong Sun\IEEEauthorrefmark{1},
Xiaojiang Du\IEEEauthorrefmark{2}, and
Mohsen Guizani\IEEEauthorrefmark{3}
}
\IEEEauthorblockA{\IEEEauthorrefmark{1}State Key Lab of ISN, Xidian University, Xi'an, 710071, China.\\ Email: \{jwliu, rsun\}@mail.xidian.edu.cn, huifangt@foxmail.com, 527690544@qq.com}
\IEEEauthorblockA{\IEEEauthorrefmark{2}Department of Computer and Information Sciences, Temple University, Philadelphia, PA 19122, USA.\\ Email: dxj@ieee.org}
\IEEEauthorblockA{\IEEEauthorrefmark{3}Department of Electrical and Computer Engineering, University of ldaho, Mosocow, ldaho, USA.\\ Email: mguizani@ieee.org}
}
\maketitle

\begin{abstract}
In recent years, cloud storage technology has been widely used in many fields such as education, business, medical and more because of its convenience and low cost. With the widespread applications of cloud storage technology, data access control methods become more and more important in cloud-based network. The ciphertext policy attribute-based encryption (CP-ABE) scheme is very suitable for access control of data in cloud storage. However, in many practical scenarios, all attributes of a user cannot be managed by one authority, so many multi-authority CP-ABE schemes have emerged. Moreover, cloud servers are usually semi-trusted, which may leak user information. Aiming at the above problems, we propose a fine-grained access control scheme with versatility for cloud storage based on multi-authority CP-ABE, named vFAC. The proposed vFAC has the features of large universe, no key escrow problem, online/offline mechanism, hidden policy, verifiability and user revocation. Finally, we demonstrate vFAC is static security under the random oracle model. Through the comparison of several existing schemes in terms of features, computational overhead and storage cost, we can draw a conclusion that vFAC is more comprehensive and scalable.
\end{abstract}
\IEEEpeerreviewmaketitle

\section{Introduction}
\begin{figure}[!t]
\setlength{\abovecaptionskip}{0pt}
\setlength{\belowcaptionskip}{10pt}
  \centering
  \includegraphics[width=8.9cm]{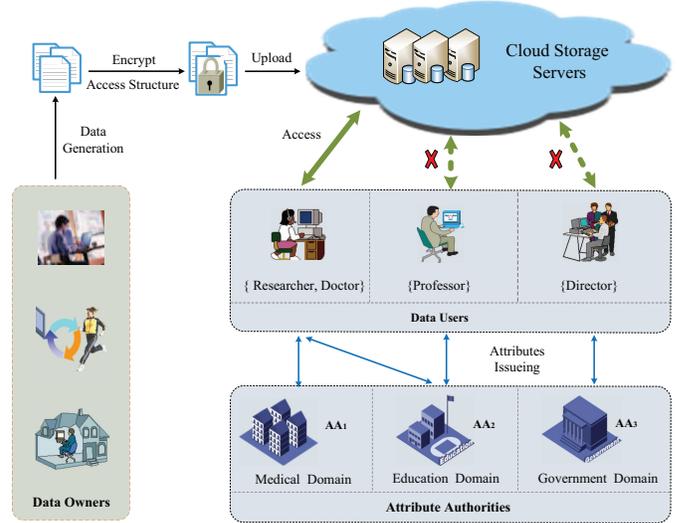}\\
  \caption{A simple architecture of data access control in cloud storage}\label{architecture}
  \vspace{-5mm}
\end{figure}

Cloud storage is an emerging network storage technology with the features of convenience and low cost. Recently, more and more users are willing to store personal data in cloud servers, in which some sensitive information might be involved\cite{zhou2013prometheus}. Therefore, data access control in cloud storage has become critical challenge. Produced by Sahai and Waters \cite{SW05} in 2005, attribute based encryption (ABE) scheme can effectively solve the data security and access control issues simultaneously. This allows users to encrypt and decrypt data based on different attributes. Following the original work, in order to provide a more complicated access control policy, CP-ABE appeared successfully. In CP-ABE, the access policy is devised by the data owner, and it is especially suitable for the designing of access control in cloud storage systems, as shown in Fig. \ref{architecture}.

With the fast development of cloud storage technology, the CP-ABE schemes with a single central authority are no longer suitable for some scenarios, because all attributes of a user are not always managed by one authority. To solve this problem, Muller et al. \cite{MKE09} proposed a multi-authority  CP-ABE system firstly in 2009, in which different attribute sets are managed by multiple authorities. Their scheme has distributed requirements by removing central authority with each attribute authority having equal status. However, most of similar schemes have the disadvantage of low efficiency. So researchers introduced online/offline mechanism and computing outsourcing technique to improve the efficiency of CP-ABE.
%These schemes \cite{GMC08,CLZ11,LBZ10,LMGS15,GHW11,LDGW13,QDLM15,LWZ15} effectively reduce the computation burden of users.
In 2008, Guo et al. \cite{GMC08} came up with an idea of identity based online/offline encryption, in which the encryption stage was split in an online phase, where only several simple operations are involved to generate the final ciphertext, and an offline phase. Since then, some schemes \cite{CLZ11,du2007effective,hei2013pipac,LDGW13,QDLM15,cheng2017lightweight,xiao2007internet} were proposed that effectively reduced the computation burden of users.

%
%Based on outsourced decryption, many ABE schemes  \cite{GHW11,LDGW13,QDLM15,LWZ15} were proposed gradually. These schemes effectively reduce the computation burden of users.

% These schemes effectively reduce the computation burden of users. In 2011, Green et al. first proposed the concept of ABE with outsourced decryption \cite{GHW11}, moving complicated calculations to a cloud server. Lai et al. \cite{LDGW13} proposed an improved ABE scheme with verifiable outsourced decryption, that allows users to verify the correctness of the partial decrypted ciphertext handled by a proxy server. Based on outsourced decryption, some ABE schemes  \cite{QDLM15,LWZ15} were proposed gradually.
%\begin{spacing}{0.4}
%%%%行间距
%\end{spacing}
%\begin{spacing}{-0.1}
%%%行间距
%\end{spacing}
Furthermore, the access policy associated with ciphertext may reveal some user sensitive information. In 2007, Kapadia et al. \cite{KTS07} protected users' privacy with hidden policy, but there were security flaws. In the next year, Nishide et al. \cite{NYO08} proposed two CP-ABE constructions to achieve hidden policy, but only partial policy was hidden. In \cite{LDL11}, a security-enhanced ABE algorithm of hidden policy was proposed in the composite order group, which proved to be completely safe under the bilinear Diffie Hellman assumption. However, the operation efficiency of bilinear pair in composite order group is lower than that of prime order group. Later, Lewko and Waters \cite{LW12} studied the security of ABE schemes in the prime order group.

Recently, there has been a lot of research on hidden policy, computational outsourcing, attribute revocation and traitor tracing according to different functional extensions. In 2015, Rouselakis et al. \cite{RW15} introduced a multi-authority ABE scheme supporting large universe, which meant that any string, as a new attribute, could be added to the system. Moreover, the number of attributes is not relevant to the public system parameters any more. In 2017, Zhang Kai et al. \cite{MZL17} solved the key escrow problem using the separate cloud server and user's private keys.
%In the same year, Yu et al. \cite{YMCZZ17} proposed a MA-ABE scheme that introduced user's accountability mechanism to solve the problem of key abuse, but this scheme involved too many attribute authorities.
At present, the latest revocation mechanisms for multi-authority ABE \cite{YMCZZ17,wu2016effective,LLL17,zhang2015toward,NMSM17,xia2016attribute} have been more flexible and can satisfy forward security, but they do not meet the feature of large universe.

In this paper, we propose a fine-grained access control scheme with versatility for cloud storage. It provides more features of online/offline mechanism, hidden policy, and verifiability than the existing schemes \cite{RW15,MZL17}. The proposed vFAC is proved to satisfy static security under the random oracle model. In addition, through performance analyses, vFAC is more comprehensive and scalable.

The rest of this paper is organized as follows: Section II reviews the related preliminaries and gives a formal definition. Section III describes the specific process of vFAC in detail. Then, section IV analyzes the security and performance through the comparison with other schemes. Finally, section V concludes the paper.

%\section{Preliminaries and formal definition}
\section{Preliminaries}
%\subsection{Preliminaries}
%\textbf{Definition 1.} Access Structure
%
%We assume that access structure ${\rm \mathbb{Z}} \subseteq {2^{\{ {K_1},{K_2}, \cdots ,{K_n}\} }}\backslash \{ \emptyset \}$ is a collection of attribute set $K = \{ {K_1},{K_2}, \cdots ,{K_n}\}$. Let the subsets in $\mathbb{Z}$ denote the authorized sets. If $\forall X,Y:X \in \mathbb{Z}$ and $X \subseteq Y$, $Y \in \mathbb{Z}$. $\mathbb{Z}$ is monotonous.

%\textbf{Definition 2.} \emph q-Decisional Parallel Bilinear Diffie-Hellman Exponent 2 (\emph {q-DPBDHE2}) Assumption
\subsection{\emph q-Decisional Parallel Bilinear Diffie-Hellman Exponent 2 (\emph {q-DPBDHE2}) Assumption}

It is a deformation based on the \emph{q-DPBDHE} assumption. We assume that $p$ is a prime number, $G$ and ${G_T}$ are multiplicative cyclic groups of order $p$, $g$ is a generator of $G$, and $e:G \times G \to {G_T}$ is a bilinear map. The following process describes the \emph{q-DPBDHE2} assumption in detail.
\begin{small}
$\begin{array}{l}
D = (p,g,G,e,{g^s},{\{ {g^{{a^i}}}\} _{i \in [2q],i \ne q + 1}},{\{ {g^{{b^j}{a^i}}}\} _{(i,j) \in [2q,q],i \ne q + 1,}}\\
{\{ {g^{\frac{s}{{{b_i}}}}}\} _{i \in [q]}},{\{ {g^{\frac{{s{a^i}{b^j}}}{{{b_{j'}}}}}}\} _{(i,j,j') \in [q + 1,q,q],j \ne j'}})
\end{array}$
\end{small}
\begin{spacing}{1.5}
%%行间距
\end{spacing}
where $a,s,{b_1}, \cdots ,{b_q} \in Z_p^ * $ are unknown, distinguishing R from $e{(g,g)^{s{a^{q + 1}}}}$ and ${G_T}$. Assuming that an attacker $\mathcal{A}$ can successfully solve the \emph{q-DPBDHE2} problem with the probability at least $\varepsilon$ in polynomial time, that is $\left| {\Pr [A(D,e{{(g,g)}^{s{a^{q + 1}}}}) = 0] - \Pr [A(D,R) = 0]} \right| \ge \varepsilon $.

It can be claimed that the advantage of solving \emph{q-DPBDHE2} problem is $\varepsilon$.
%If \emph{(q-DPBDHE2)} problem is hard, then $\varepsilon$ can be ignorable.
\begin{spacing}{0.5}
%%行间距
\end{spacing}
\subsection{Formal Definition}
Let $U$ represent attribute space, and each attribute authority $A{A_i}(i \in \left[ {1,n} \right])$  manages its own attribute domain ${U_i} \in U$. For $\forall k,l \in \left[ {1,n} \right]$, $k \ne l$ , then ${U_k} \cap {U_l} = \emptyset$. This scheme contains eight formal algorithms.

\textbf{GlobalSetup}$(\lambda ) \to GP$: The $GlobalSetup$ algorithm inputs the security parameter $\lambda$  and outputs global parameters $GP$.

\textbf{AuthoritySetup}$(GP,i) \rightarrow \langle {\rm{P}}{{\rm{K}}_i}, {\rm{S}}{{\rm{K}}_i}  \rangle$: This algorithm only inputs $GP$ and attribute authority $i$, and generates its public/secret key pair $ \langle {\rm{P}}{{\rm{K}}_i}, {\rm{S}}{{\rm{K}}_i}  \rangle$.

%\textbf{CSKeyGen}\begin{small}$(GP,\left\{ {{\rm{S}}{{\rm{K}}_i}} \right\},GID,UP{K_{GID}},S)$\end{small} $\to  CS{K_{GID,S}}$:
%The $CSKeyGen$ algorithm inputs $GP$, user's $GID$ and public key $UP{K_{GID}}$ , a set of user's attributes $S$ and secret key $\left\{ {{\rm{S}}{{\rm{K}}_i}} \right\}$ of the relevant attribute authorities. It generates the private key $CS{K_{GID,S}}$ of the corresponding cloud server, then adds the private key and corresponding identifier to the key list $KT$.
%
%\textbf{UserKeyGen}$(GP,GID) \to \langle UP{K_{GID}},US{K_{GID}}\rangle $: The $UserKeyGen$ algorithm inputs $GP$ and user's global identifier $GID$, then outputs user's public/secret key pair $ \langle UP{K_{GID}},US{K_{GID}}\rangle $.

\textbf{KeyGen}\begin{small}$(GP,GID,\left\{ {{\rm{S}}{{\rm{K}}_i}} \right\},S)  \to \langle UP{K_{GID}},  CS{K_{GID,S}},\\
 US{K_{GID}}\rangle$:\end{small}
The $KeyGen$ algorithm inputs $GP$, user's $GID$, secret key $\left\{ {{\rm{S}}{{\rm{K}}_i}} \right\}$ of the relevant attribute authorities and a set of the user's attributes $S$. It outputs user's public key $UP{K_{GID}}$, the private key $CS{K_{GID,S}}$ of the corresponding cloud server and user's secret key $US{K_{GID}} $.

\textbf{Offline.Enc}$(GP,\left\{ {{\rm{P}}{{\rm{K}}_i}} \right\}) \to IC$: The $Offline.Enc$ algorithm inputs $GP$ and outputs intermediate ciphertext $IC$.

\textbf{Online.Enc}$(GP,\left\{ {{\rm{P}}{{\rm{K}}_i}} \right\},M,IC,A) \to CT$: This algorithm inputs $GP$, message $M$, intermediate ciphertext $IC$, access policy $A$ and public key $\left\{ {{\rm{P}}{{\rm{K}}_i}} \right\}$ of the relevant attribute authorities. It outputs ciphertext $CT$.

\textbf{CS.Dec}$(GP,CS{K_{GID,S}},UP{K_{GID}},CT) \to C{T_{GID}}$ or $ \bot $: The $CS.Dec$ algorithm inputs $GP$, secret key $CS{K_{GID,S}}$ of cloud server, public key $UP{K_{GID}}$  of the user, and ciphertext $CT$. Then, it outputs partial decrypted ciphertext $C{T_{GID}}$ or a symbol $ \bot $ which represents ciphertext cannot be decrypted successfully.

\textbf{User.Dec}$(US{K_{GID}},C{T_{GID}}) \to M$ or $ \bot $: The $User.Dec$ algorithm inputs user's public key $US{K_{GID}}$ and partial decrypted ciphertext $C{T_{GID}}$. It outputs the recovered message $M$ or $ \bot $.

\textbf{Revoke}$(GID,KT) \to KT/\{ GID,CS{K_{GID}}\}$: This algorithm inputs a user's $GID$ and a key list $KT$, and outputs the key list $KT$ after revocation.

\section{Fine-Grained Access Control with Versatility for Cloud Storage}
\subsection{System Model}

\begin{figure}[!tb]
\setlength{\abovecaptionskip}{0pt}
\setlength{\belowcaptionskip}{10pt}
  \centering
  \includegraphics[width=8.6cm]{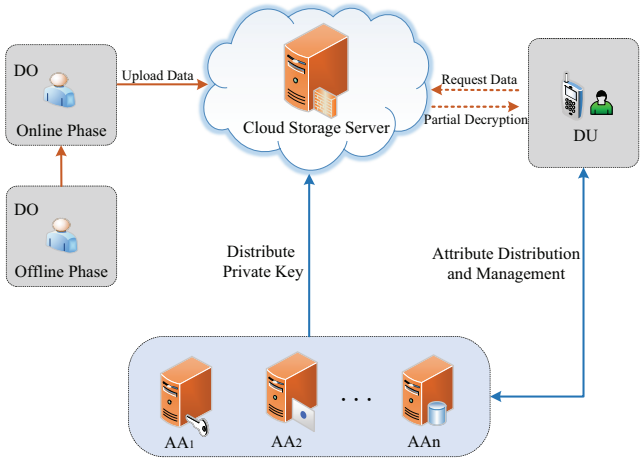}\\
  \caption{System model}\label{model}
  %\vspace{-1mm}
\end{figure}

In Fig.\ref{model}, we can see that the system contains four participants: Attribute Authority (AA), Cloud Server (CS), Data Owner (DO), and Data User (DU).
%\begin{spacing}{0.4}
%%%%行间距
%\end{spacing}
%\begin{spacing}{-0.2}
%%行间距
%\end{spacing}

AA: It is in charge of managing the DU's attribute set, and generating the corresponding CS's private key for these attributes.

CS: It stores encrypted data and manages the CS's private keys corresponding to users.

DO: DO encrypts data based on the access policy, then uploads the encrypted data to CS.

DU: DU can request data from CS. If the attributes of DU satisfy the access structure, CS will return the corresponding partial decrypted ciphertext, then DU restores the cipher with his/her own private key.

\subsection{Security Model}
First, we define a static security model which requires query-response phase to be completed before the challenge phase. During the query phase, an attacker $\mathcal{A}$ can query the private key of DU and CS, and control some attribute authorities.
The specific description is as follows:

\textbf{Setup: } A challenger $\mathcal{C}$ generates $GP$ by $GlobalSetup$ algorithm and sends it to $\mathcal{A}$.

\textbf{Query-response Phase: }Assume ${U_\theta }$ is the set of attribute authorities, ${C_\theta }$ is the set of partial attribute authorities controlled by $\mathcal{A}$, and ${N_\theta }$ is the set of other attribute authorities that are not controlled by $\mathcal{A}$ .
\begin{itemize}
\item
$\mathcal{A}$ submits an uncontrolled attribute authority $\theta  \in {N_\theta }$, then $\mathcal{C}$ runs the $AuthoritySetup$ algorithm and returns the public key $P{K_\theta }$ of $\theta $.
\item
$\mathcal{A}$ submits the DU's global identifier $GI{D_i}$, then $\mathcal{C}$ executes the $KeyGen$ algorithm and returns the DU's public and private key pair $\left\langle {P{K_i},S{K_i}} \right\rangle$.
\item
$\mathcal{A}$ submits the DU's global identifier $GI{D_i}$ and the corresponding attribute set ${S_i}$, then $\mathcal{C}$ executes the $KeyGen$ algorithm and returns the private key $CS{K_{GI{D_i},{S_i}}}$ of the CS.

 %There is no need to query if an attribute in the attribute set ${S_i}$ is managed by one of the attribute authorities of ${N_\theta }$. Because $\mathcal{A}$ can control this authority to generate the corresponding private key of cloud server. In addition, in this inquiry, $\mathcal{A}$ can both query the cloud server's private keys that correspond to the users who have already inquired about the public and private key pair, and can also query the cloud server's private keys corresponding to other users.
\end{itemize}

\textbf{Challenge: }$\mathcal{A}$ submits the challenge access structure $\left( {{A^ * },{\rho ^ * }} \right)$ and the challenge ciphertext $M_0^ * $, $M_1^ * $. $\mathcal{C}$ randomly selects $b \in \{ 0,1\}$, and executes $Offline.Enc$ and $Online.Enc$ algorithms in turns and returns the challenge ciphertext $C{T^ * }$. Note that for any user $GI{D_i}$ who has queried for a private key, the attribute set ${S_{{C_\theta }}} \cup {S_i}$ cannot satisfy the challenge access structure $({A^ * },{\rho ^ * })$.

\textbf{Guess: } $\mathcal{A}$ outputs a bit $b' \in \{ 0,1\}$.

The attacker's winning advantage can be defined as $\left| {\Pr \left[ {b = b'} \right] - \frac{1}{2}} \right|$.

\subsection{Our Scheme}
Based on the system model and formal definition, vFAC is described as follows.

%\subsubsection{System Initialization}\
\emph{1) System Initialization}

\textbf{GlobalSetup:} In this algorithm, a bilinear map ${\rm{e}}:G \times G \to {G_T}$ is chosen firstly, where the orders of $G$ and ${G_T}$ are both large prime number $p$, and $g$ is a generator of $G$. Next, select a symmetric algorithm $SE = (SE.Enc,SE.Dec,{l_{SE}})$, where $SE.Enc$ is the encryption algorithm, $SE.Dec$ is the decryption algorithm, and ${l_{SE}}$ represents the length of the secret key. Then, choose five strong collision-resistant hash functions: \begin{small} $H:Z_P^ *  \to G,$ $F:U \to G,$ $h:{G_T} \to {\{ 0,1\} ^{{l_{SE}}}},$ ${H_1}:{G_T} \to {\{ 0,1\} ^{{l_{H1}}}},$ ${H_2}:{\{ 0,1\} ^ * } \to {\{ 0,1\} ^{{l_{H2}}}}$. \end{small} Finally, publish global parameters $GP$: $GP =\langle\lambda ,e,G,{G_T},p,g,U,\{ {U_i}\} ,H,F,h,{H_1},{H_2},SE\rangle $.

\textbf{AuthoritySetup:} Each attribute authority $i \in [1,n]$ randomly selects ${\alpha _i},{\beta _i},{y_i} \in Z_p^ * $, then sets its own secret key as $S{K_i} = \langle{\alpha _i},{\beta _i},{y_i}\rangle $ and public key as $P{K_i}=\langle e{(g,g)^{{\alpha _i}}},{g^{{\beta _i}}},{g^{{y_i}}}\rangle$.

%\subsubsection{Key Generation}\
\emph{2) Key Generation}

\textbf{KeyGen:} The user $GID$ chooses a random number ${x_{GID}} \in Z_p^ * $, then sets his/her public key as $UP{K_{GID}}=\langle{g^{{x_{GID}}}},H{(GID)^{{x_{GID}}}}\rangle$. For each attribute $j \in S$, if it is managed by the attribute authority $i$, $i$ needs to choose ${t_j} \in Z_p^ * $ randomly, calculate \begin{small}$K_{j,GID}^1 = {g^{{x_{GID}}{\alpha _i}}}H{(GID)^{{x_{GID}}{y_i}}}F{(j)^{{t_j}}},$ $K_{j,GID}^2 = {g^{{t_j}}} ,$ $K_{j,GID}^3 = F{(j)^{{\beta _i}}}$\end{small}, and set the CS's private key corresponding to the $GID$ as $CS{K_{GID,S}} = {\{ K_{j,GID}^1,K_{j,GID}^2\} _{j \in S}}$. Then, the attribute authority $i$ adds $\langle GID,CS{K_{GID,S}}\rangle$ to the key list $KT$ and sends ${\left\{ {{\rm{K}}_{j,GID}^3} \right\}_{j \in S}}$ to the user $GID$ through a secure channel.

On receiving the ${\left\{ {{\rm{K}}_{j,GID}^3} \right\}_{j \in S}}$, the user $GID$ sets his/her secret key as $US{K_{GID}} = \{ x_{GID}^{ - 1},{\{ K_{j,GID}^3\} _{j \in S}}\}$.

\emph{3) Offline/Online Data Encryption}

\textbf{Offline.Enc:} For each attribute $j \in [1,U]$, DO randomly selects ${\lambda _j}^\prime ,{r_j},{w_j}^\prime  \in Z_p^ * $, precomputes the ciphertext ${C_{1,j}} = e{(g,g)^{{\lambda _j}^\prime }}e{(g,g)^{{\alpha _{\delta (j)}}{r_j}}},$ ${C_{2,j}} = {g^{ - {r_j}}},$ ${C_{3,j}} = {g^{{y_{\delta (j)}}{r_j}}}{g^{{w_j}^\prime }},$ ${C_{4,j}} = F{(j)^{{r_j}}}$, and outputs the intermediate ciphertext: $IC = {\{ {\lambda _j}^\prime ,{w_j}^\prime ,{C_{1,j}},{C_{2,j}},{C_{3,j}},{C_{4,j}}\} _{j \in [1,U]}}$.

\textbf{Online.Enc:} Suppose that DO's attribute domain for creating access policy is $D$. In this phase, DO randomly selects $a \in Z_p^ * $, calculates ${\sigma _j} = e({({g^{{\beta _{\delta (j)}}}})^a},F(j))$ for each attribute $j \in D$, and replaces $j$ with ${H_1}({\sigma _j})$, where $\delta (j)$ represents the authority who manages the attribute $j$. DO uses the replaced attributes to generate the access policy $(A,\rho )$, where A is a $l \times n$ matrix and $\rho $ is a map from the row of matrix A to $D$. Then, DO generates the ciphertext by doing the following:

\begin{itemize}
\item
Randomly select $s,{y_2}, \cdots ,{y_n},{z_2}, \cdots ,{z_n} \in Z_p^ * $ and build vectors $\vec v  = {(s,{y_2}, \cdots ,{y_n})^T}$, $\vec w  = {(0,{z_2}, \cdots ,{z_n})^T}$.
\item
Compute ${\lambda _j} = \vec A_{\rm{j}}  \cdot {\vec v} $, ${w_j} = \vec {{A_j}}  \cdot \vec w $, where ${\vec A_{\rm{j}}}$ represents the row vector in the matrix A that corresponds to $j$.
\item
Randomly select $M,R \in {G_T}$, and compute $h = {g^a}$, ${C_0} = {\mathop{\rm Re}\nolimits} {(g,g)^s}$, ${C_{5,j}} = {\lambda _j} - {\lambda _j}^\prime $, ${C_{6,j}} = {w_j} - {w_j}^\prime$, ，${K_{SE}} = h(R)$, ${C_{SE}} = SE.Enc({K_{SE}},M)$, $Tag = {H_1}(R)$, $V{K_M} = {H_2}(Tag\parallel {C_{SE}})$.
\end{itemize}

Finally, the ciphertext $CT$ is uploaded to the CS.
\begin{small}
\begin{center}
$\begin{array}{l}
CT = \langle(A,\rho ),{C_0},h,{C_{SE}},V{K_M},\{ {C_{1,j}},{C_{2,j}},{C_{3,j}},{C_{4,j}},{C_{5,j}},\\
{C_{6,j}}{\} _{j \in D}}\rangle.
\end{array} $
\end{center}
\end{small}
%subsubsection{ Cloud Server Data Decryption}\
\emph{4)  Data Decryption}

\textbf{CS.Dec:} When DU requests the CS to decrypt the ciphertext $CT$, s/he first downloads $h$ securely from $CT$, then computes ${\sigma _j} = e(h,K_{j,GID}^3)$ for each attribute $j$ and replaces $j$ with ${H_1}({\sigma _j})$. If ${H_1}({\sigma _j})$ satisfies the access structure $(A,\rho )$, the CS must be able to find a set of constants $\{ {c_j} \in {Z_p}\} $ to make it satisfy
\begin{small}$\sum\limits_{j \in I \subseteq \{ 1,2, \cdots l\} } {{c_j}\vec {{A_j}} = (1,0, \cdots ,0)} $\end{small}. Next, the CS calculates
\begin{small}
$\textstyle {C_{1,GID}} = {\prod\limits_{j \in I} {({C_{1,j}}e{{(g,g)}^{{C_{5,j}}}})} ^{{c_j}}}$
\end{small},
\begin{small}
${C_{2,GID}} = {\prod\limits_{j \in I} {(e(K_{j,GID}^1,{C_{2,j}})e(H{{(GID)}^{{x_{GID}}}},{C_{3,j}}{g^{{C_{6,j}}}})e(K_{j,GID}^2,{C_{4,j}}))} ^{{c_j}}}$
\end{small}
 and returns the partial decrypted ciphertext $C{T_{GID}} = \langle {C_0},{C_{1,GID}},{C_{2,GID}},V{K_M},{C_{SE}}\rangle $. Otherwise, CS returns $ \bot $ to DU if ${H_1}({\sigma _j})$
  does not satisfy the access structure $(A,\rho )$.
%\subsubsection{ User Data Decryption}\

\textbf{User.Dec:} Upon receiving $C{T_{GID}}$, DU calculates ${C_{1,GID}}{C_{2,GID}}^{x_{GID}^{ - 1}} = e{(g,g)^s}$, $R = \frac{{{C_0}}}{{e{{(g,g)}^s}}}$, $Tag = {H_1}(R)$. Then, DO verifies if the equation ${H_2}(Tag\parallel {C_{SE}}) = V{K_M}$ holds. If it does, DU continues to calculate ${K_{SE}} = h(R)$, $M = SE.Dec({K_{SE}},{C_{SE}})$, and returns $M$. Otherwise, it returns $\bot $.

\emph{5)  User Revocation}

\textbf{Revoke:} To revoke the user $GID$, the CS can find the corresponding entry from the key list and delete it.
\begin{table*}[htbp]
\scriptsize
\caption{Comparison of Features}\label{tabl:comparison}
\begin{center}
\begin{tabular}{c p{2cm}<{\centering} c c c  c c p{1cm}<{\centering} c}
%\toprule
  \hline \hline

  \raisebox{-0.06cm}[0pt]{\textbf{Schemes}} & \raisebox{-0.06cm}[0pt]{ \textbf{Prime order group} } & \raisebox{-0.06cm}[0pt]{\textbf{No CA }} & \raisebox{-0.06cm}[0pt]{\textbf{Large universe} } & \raisebox{-0.06cm}[0pt]{\textbf{No key escrow problem} } & \raisebox{-0.06cm}[0pt]{\textbf{Online/Offline }} & \raisebox{-0.06cm}[0pt]{\textbf{Hidden policy }} &\raisebox{-0.06cm}[0pt]{\textbf{Verifiability}}&
  \raisebox{-0.06cm}[0pt]{\textbf{Revocation}} \\
%\toprule
 \hline \hline
  \raisebox{-0.05cm}[0pt]{RW\cite{RW15}} & \raisebox{-0.05cm}[0pt]{${\surd}$} & \raisebox{-0.05cm}[0pt]{${\surd}$} & \raisebox{-0.05cm}[0pt]{${\surd}$} & \raisebox{-0.05cm}[0pt]{$ \times $} & \raisebox{-0.05cm}[0pt]{$ \times $} & \raisebox{-0.05cm}[0pt]{$ \times $} & \raisebox{-0.05cm}[0pt]{$ \times $} & \raisebox{-0.05cm}[0pt]{$ \times $}\\

  %\hline
  \raisebox{-0.05cm}[0pt]{MZL\cite{MZL17}} & \raisebox{-0.05cm}[0pt]{${\surd}$} & \raisebox{-0.05cm}[0pt]{${\surd}$} & \raisebox{-0.05cm}[0pt]{${\surd}$} & \raisebox{-0.05cm}[0pt]{${\surd}$} & \raisebox{-0.05cm}[0pt]{$ \times $} & \raisebox{-0.05cm}[0pt]{$ \times $} & \raisebox{-0.05cm}[0pt]{$ \times $} &
  \raisebox{-0.05cm}[0pt]{${\surd}$}\\

 % \hline
  \raisebox{-0.05cm}[0pt]{YMCZZ\cite{YMCZZ17}} & \raisebox{-0.05cm}[0pt]{$ \times $} & \raisebox{-0.05cm}[0pt]{${\surd}$} & \raisebox{-0.05cm}[0pt]{$ \times $} & \raisebox{-0.05cm}[0pt]{${\surd}$} & \raisebox{-0.05cm}[0pt]{$ \times $} &
  \raisebox{-0.05cm}[0pt]{$ \times $} & \raisebox{-0.05cm}[0pt]{$ \times $} & \raisebox{-0.05cm}[0pt]{$ \times $}\\

  \raisebox{-0.05cm}[0pt]{LLL\cite{LLL17}} & \raisebox{-0.05cm}[0pt]{${\surd}$} & \raisebox{-0.05cm}[0pt]{${\surd}$} & \raisebox{-0.05cm}[0pt]{$ \times $} & \raisebox{-0.05cm}[0pt]{${\surd}$} & \raisebox{-0.05cm}[0pt]{${\surd}$} &
  \raisebox{-0.05cm}[0pt]{${\surd}$} & \raisebox{-0.05cm}[0pt]{${\surd}$} & \raisebox{-0.05cm}[0pt]{${\surd}$}\\

  \raisebox{-0.05cm}[0pt]{NMSM\cite{NMSM17}} & \raisebox{-0.05cm}[0pt]{${\surd}$} & \raisebox{-0.05cm}[0pt]{$ \times $} & \raisebox{-0.05cm}[0pt]{$ \times $} & \raisebox{-0.05cm}[0pt]{${\surd}$} & \raisebox{-0.05cm}[0pt]{$ \times $} &
  \raisebox{-0.05cm}[0pt]{$ \times $} & \raisebox{-0.05cm}[0pt]{$ \times $} & \raisebox{-0.05cm}[0pt]{${\surd}$}\\

  %\raisebox{-0.05cm}[0pt]{[35]} & \raisebox{-0.05cm}[0pt]{$ \times $} & \raisebox{-0.05cm}[0pt]{$ \times $} & \raisebox{-0.05cm}[0pt]{$ \times $} & \raisebox{-0.05cm}[0pt]{$ \times $} & \raisebox{-0.05cm}[0pt]{$ \times $} &
%  \raisebox{-0.05cm}[0pt]{$ \times $} & \raisebox{-0.05cm}[0pt]{$ \times $} \\

  \raisebox{-0.05cm}[0pt]{vFAC} & \raisebox{-0.05cm}[0pt]{${\surd}$} & \raisebox{-0.05cm}[0pt]{${\surd}$} & \raisebox{-0.05cm}[0pt]{${\surd}$} & \raisebox{-0.05cm}[0pt]{${\surd}$} & \raisebox{-0.05cm}[0pt]{${\surd}$} &
  \raisebox{-0.05cm}[0pt]{${\surd}$} & \raisebox{-0.05cm}[0pt]{${\surd}$} & \raisebox{-0.05cm}[0pt]{${\surd}$}\\

%\toprule
  \hline \hline

\end{tabular}
\end{center}
\end{table*}
\vspace{0.4em}

\section{SECURITY AND PERFORMANCE ANALYSES}
%Here, we first verify the correctness of vFAC. Then, the security analysis is given. Finally, the features and cost of calculation and storage are compared among several selected schemes.
\subsection{Correctness Analysis}
%We can demonstrate that our vFAC is correct by the following formulas.
If a DU's attributes satisfy the access structure, the equations $\sum\limits_{j \in I} {{\lambda _j}{c_j}}  = s$ and $\sum\limits_{j \in I} {{w_j}{c_j}}  = 0$ will hold. Then, we can have the following formulas:
%If a DU's attributes satisfy the access structure, the equations (1) and (2) will hold.
%\begin{spacing}{0.6}
%\end{spacing}
% %\begin{small}
% \begin{align}
% \sum\limits_{j \in I} {{\lambda _j}{c_j}}  = s
% \end{align}
%%\end{small}
% %\begin{spacing}{-0.5}
%%%%行间距
%%\end{spacing}
% %\begin{small}
% \begin{align}
% \sum\limits_{j \in I} {{w_j}{c_j}}  = 0
%  \end{align}
% \end{small}

%Based on the above equations, we can have the following formulas:
%\begin{equation}
%\begin{aligned}
%{\sigma _j}= e({({g^{{\beta _{\delta (j)}}}})^a},F(j))= e({g^a},F{(j)^{^{{\beta _{\delta (j)}}}}})= e(h,K_{j,GID}^3)
%\end{aligned}
%\end{equation}
%
%\begin{spacing}{0.6}
%\end{spacing}
%\begin{equation}
%\begin{aligned}
%{C_{1,j}}e{(g,g)^{{C_{5,j}}}} &= e{(g,g)^{{\lambda _j}^\prime }}e{(g,g)^{{\alpha _{\delta (j)}}{r_j}}}e{(g,g)^{{\lambda _j} - {\lambda _j}^\prime }}\\
%&= e{(g,g)^{{\lambda _j}}}e{(g,g)^{{\alpha _{\delta (j)}}{r_j}}}\\
%\end{aligned}
%\end{equation}
%%\begin{spacing}{0.5}
%%\end{spacing}
%\begin{equation}
%
%\begin{align}
%{C_{3,j}}{g^{{C_{6,j}}}} = {g^{{y_{\delta (j)}}{r_j}}}{g^{{w_j}^\prime }}{g^{{w_j} - {w_j}^\prime }}= {g^{{y_{\delta (j)}}{r_j}}}{g^{{w_j}}}
%\end{align}
%
%\end{equation}
\begin{small}
\begin{equation}
\begin{aligned}
{\sigma _j} &= e({({g^{{\beta _{\delta (j)}}}})^a},F(j))= e({g^a},F{(j)^{^{{\beta _{\delta (j)}}}}})= e(h,K_{j,GID}^3)
\end{aligned}
\end{equation}
\end{small}
\begin{spacing}{-1}
\end{spacing}
\begin{equation}
\begin{aligned}
{C_{1,j}}e{(g,g)^{{C_{5,j}}}} &= e{(g,g)^{{\lambda _j}^\prime }}e{(g,g)^{{\alpha _{\delta (j)}}{r_j}}}e{(g,g)^{{\lambda _j} - {\lambda _j}^\prime }}\\
&= e{(g,g)^{{\lambda _j}}}e{(g,g)^{{\alpha _{\delta (j)}}{r_j}}}\\
\end{aligned}
\end{equation}
\begin{spacing}{0.5}
\end{spacing}
\begin{equation}
\begin{aligned}
{C_{3,j}}{g^{{C_{6,j}}}} &= {g^{{y_{\delta (j)}}{r_j}}}{g^{{w_j}^\prime }}{g^{{w_j} - {w_j}^\prime }}= {g^{{y_{\delta (j)}}{r_j}}}{g^{{w_j}}}
\end{aligned}
\end{equation}
\begin{spacing}{0.5}
\end{spacing}
\begin{small}
\begin{align}
\begin{array}{l}
{C_{1,GID}}{C_{2,GID}}^{x_{GID}^{ - 1}}\\
 = {\prod\limits_{j \in I} {({C_{1,j}}e{{(g,g)}^{{C_{5,j}}}})} ^{{c_j}}}\prod\limits_{j \in I} {(e(K_{j,GID}^1,{C_{2,j}}) \cdot } \\
e(H{(GID)^{{x_{GID}}}},{C_{3,j}}{g^{{C_{6,j}}}})e(K_{j,GID}^2,{C_{4,j}}){)^{\frac{{{c_j}}}{{{x_{GID}}}}}}\\
 = \prod\limits_{j \in I} \{  e{(g,g)^{{\lambda _j}}}e{(g,g)^{{\alpha _{\delta (j)}}{r_j}}} \cdot \\
e{({g^{^{{x_{GID}}{\alpha _{_{\delta (j)}}}}}}H{(GID)^{{x_{GID}}{y_{\delta (j)}}}}F{(j)^{{t_j}}},{g^{ - {r_j}}})^{\frac{1}{{{x_{GID}}}}}} \cdot \\
e{(H{(GID)^{{x_{GID}}}},{g^{{y_{\delta (j)}}{r_j}}}{g^{{w_j}}})^{^{\frac{1}{{{x_{GID}}}}}}}e{({g^{{t_j}}},F{(j)^{^{{r_j}}}})^{\frac{1}{{{x_{GID}}}}}}{\} ^{{c_j}}}\\
 = \prod\limits_{j \in I} \{  e{(g,g)^{{\lambda _j}}}e{(H(GID),g)^{{w_j}}}e({g^{{\alpha _{\delta (j)}}}},{g^{ - {r_j}}}) \cdot \\
e(H{(GID)^{^{{y_{\delta (j)}}}}},{g^{ - {r_j}}})e(F{(j)^{\frac{{{t_j}}}{{{x_{GID}}}}}},{g^{ - {r_j}}})e({g^{\frac{{{t_j}}}{{{x_{GID}}}}}},F{(j)^{{r_j}}}) \cdot \\
e(H(GID),{g^{{y_{\delta (j)}}{r_j}}})e{(g,g)^{^{{\alpha _{\delta (j)}}{r_j}}}}{\} ^{{c_j}}}\\
 = \prod\limits_{j \in I} \{  e{(g,g)^{{\lambda _j}}}e{(H(GID),g)^{{w_j}}}{\} ^{{c_j}}}\\
 = e{(g,g)^{\sum\limits_{j \in I} {{\lambda _j}{c_j}} }}e{(H(GID),g)^{\sum\limits_{j \in I} {{w_j}{c_j}} }}\\
 = e{(g,g)^s}
\end{array}
\end{align}
\end{small}

If we can restore $e{(g,g)^s}$, the plaintext $M$ will be decrypted correctly.
\subsection{Security Analysis}
In this subsection, we analyze the security properties of the vFAC in the following respects.

\subsubsection{Static Security}
Here, we analyze the security of vFAC based on security model in Section III.

\newtheorem{lemma}{Lemma}
\begin{lemma}
If the scheme in \cite{RW15}, named RW, satisfies the static security under the random oracle model, vFAC can also satisfy the static security.
\end{lemma}

\textbf{\emph{Proof.}}
Assume that, under the static security model, an attacker $\mathcal{A}$ can break vFAC in polynomial time by the advantage $\varepsilon $. So, there must be a simulator $\mathcal{B}$ can break RW with the same advantage. The following specifically describes how a simulator $\mathcal{B}$ breaks RW with the help of $\mathcal{A}$ and the challenger $\mathcal{C}$ of RW.

\textbf{Setup. } $\mathcal{C}$ executes $GlobalSetup$ algorithm in RW and sends $GP$ to $\mathcal{B}$. According to the $GlobalSetup$ algorithm of vFAC, $\mathcal{B}$ generates the global parameters $GP$ and sends it to $\mathcal{A}$.

\textbf{Query-response Phase. }In this phase, we assume that the set of attribute authorities is ${U_\theta }$, the set of corrupted authorities controlled by $\mathcal{A}$ is ${C_\theta }$, and the set of uncontrolled authorities is ${N_\theta }$, besides, ${N_\theta } \cup {C_\theta } = {U_\theta }$, ${N_\theta } \cap {C_\theta } = \emptyset $. For a corrupted attribute authority $\theta  \in {C_\theta }$, $\mathcal{A}$ first generates the corresponding public key ${\{ P{K_\theta }\} _{\theta  \in {C_\theta }}}$ of $\theta $ and sends it to $\mathcal{B}$. Then, $\mathcal{B}$ sends ${\{ P{K_\theta }\} _{\theta  \in {C_\theta }}}$ to $\mathcal{C}$. Next, $\mathcal{A}$ does the following queries to $\mathcal{B}$, and $\mathcal{B}$ gives the corresponding responses.
\begin{itemize}
\item
$\mathcal{A}$ submits an uncontrolled attribute authority $\theta  \in {N_\theta }$ , then $\mathcal{B}$ asks $\mathcal{C}$ for the corresponding public key of $\theta $. $\mathcal{C}$ executes the $AuthoritySetup$ algorithm of RW, generates the corresponding public key $P{K_\theta } = \langle e{(g,g)^{{\alpha _\theta }}},{g^{{y_\theta }}} \rangle $, and sends it to $\mathcal{B}$. Then $\mathcal{B}$ updates the public key to $P{K_\theta } = \langle e{(g,g)^{{\alpha _\theta }}},{g^{{\beta _\theta }}},{g^{{y_\theta }}}\rangle $ and sends $P{K_\theta }$ to $\mathcal{A}$ according to the $AuthoritySetup$ algorithm of vFAC.
\item
$\mathcal{A}$ submits a user's identifier $GI{D_i}(1 \le i \le m)$ to $\mathcal{B}$, then $\mathcal{B}$ executes the $KeyGen$ algorithm to generate the corresponding private key $US{K_{GI{D_i}}} = x_{GI{D_i}}^{ - 1}$, public key $UP{K_{GI{D_i}}} = \langle {g^{{x_{GI{D_i}}}}},H{(GI{D_i})^{{x_{GI{D_i}}}}} \rangle $, and sends $ \langle US{K_{GI{D_i}}},UP{K_{GI{D_i}}} \rangle $ to $\mathcal{A}$.
\item
$\mathcal{A}$ submits a user's identifier $GI{D_i}$ and the user's attribute set ${S_{\rm{i}}}\left( {1 \le i \le {\rm{n}},m < n} \right)$ to $\mathcal{B}$, then $\mathcal{B}$ returns the corresponding CS's private key and user's private key to $\mathcal{A}$. If $1 \le i \le {\rm{m}}$, then for each $j \in {S_{\rm{i}}}$, $\mathcal{B}$ chooses ${t_j} \in Z_p^ * $ randomly, and computes \begin{small}$K_{j,GI{D_i}}^1 = {g^{{x_{GI{D_i}}}{\alpha _\theta }}}H{(GI{D_i})^{{x_{GI{D_i}}}{y_\theta }}}F{(j)^{{t_j}{x_{GI{D_i}}}}}$, $K_{j,GI{D_i}}^2 = {g^{{t_j}{x_{GI{D_i}}}}}$, $K_{j,GI{D_i}}^3 = F{(j)^{{\beta _\theta }}}$\end{small}; If $m < i \le {\rm{n}}$, then $\mathcal{B}$ chooses ${g_j} \in G,{t_j} \in Z_p^ * $ randomly, and computes \begin{small}$K_{j,GI{D_i}}^1 = {g_j}F{(j)^{{t_j}}}$, $K_{j,GI{D_i}}^2 = {g^{{t_j}}}$, $K_{j,GI{D_i}}^3 = F{(j)^{{\beta _\theta }}}$\end{small}.
%Because in the cyclic group G, there must be an unknown private key ${x_{GI{D_i}}} \in Z_p^ * $, so that ${g_j} = {\left( {{g^{{\alpha _\theta }}}H{{\left( {GI{D_i}} \right)}^{{y_\theta }}}} \right)^{{x_{GI{D_i}}}}}$. Therefore, in this case, $K_{j,GI{D_i}}^1 = {g_j}F{\left( j \right)^{{t_j}}} = {g^{{\alpha _\theta }{x_{GI{D_i}}}}}H{(GI{D_i})^{{y_\theta }{x_{GI{D_i}}}}}F{(j)^{{t_j}}}$ and $K_{j,GI{D_i}}^2$ are also in the form of the private key of the cloud server in our scheme.
Finally, $\mathcal{B}$ sends $\mathcal{A}$ the corresponding private keys of CS and user.
\end{itemize}

\textbf{Challenge. }$\mathcal{A}$ submits the challenge access structure $\left( {{A^ * },{\rho ^ * }} \right)$, challenge plaintext  $M_0^ * $, $M_1^ * $ to $\mathcal{B}$. $\mathcal{B}$ randomly selects $b \in \{ 0,1\} $, executes $Offline.Enc$ and $Online.Enc$ algorithms, and returns the challenge ciphertext $C{T^ * }$ to $\mathcal{A}$. Note that for all users who have queried the private key, and the attribute set ${S_{{C_\theta }}} \cup {S_i}$ cannot satisfy the challenge access structure $({A^ * },{\rho ^ * })$.

\textbf{Guess. }$\mathcal{A}$ outputs a bit $b' \in \{ 0,1\}$, $\mathcal{B}$ also outputs $b'$.

In the above game, $\mathcal{B}$ perfectly simulates the challenger of vFAC under real conditions, and the CS's private key generated by $\mathcal{B}$ matches the user's private key generated by the $KeyGen$ algorithm of RW. In addition, $\mathcal{B}$ can determine the selected R in the $Online.Enc$ algorithm of vFAC from the $b'$.

R is equivalent to the message $M$ that needs to be encrypted in the $Encrypt$ algorithm of RW. Therefore, if the attacker $\mathcal{A}$ can break vFAC with the advantage $\varepsilon $ in polynomial time, $\mathcal{B}$ can also break RW, which contradicts with the premise that RW satisfies static security.

\newtheorem{theorem2}{Theorem2}
\begin{lemma}
If the q-DPBDHE2 assumption holds, RW satisfies the static security under the random oracle model.
\end{lemma}

\textbf{\emph{Proof.}} Lemma 2 has been proven in \cite{RW15}.

\newtheorem{theorem}{Theorem}
\begin{theorem}
Our proposed vFAC satisfies the static security under the random oracle model.
\end{theorem}
\textbf{\emph{Proof.}} According to Lemma 1 and Lemma 2, Theorem 1 can be proven.

\subsubsection{Hidden Policy}
In the $online.Enc$ phase, $DO$ replaces all attributes of $D$ with ${H_1}({\sigma _j})$, and only the user $GI{D_i}$ with the corresponding key $K_{j,GI{D_i}}^3$ can recover ${\sigma _j}$ for each attribute $j \in D$. The access policy of the ciphertext stored on the CS does not provide any useful information about user attributes, so privacy protection for user attributes can be achieved.
\subsubsection{No Key Escrow Problem}
DU's public key $UP{K_{GI{D_i}}}$ is used as a generating parameter when AA generates the corresponding private key of CS for the user $GI{D_i}$. Therefore, whether AA or CS can only partially decrypt the ciphertext, and only the user can restore the corresponding plaintext with his/her private key $x_{GI{D_i}}^{ - 1}$. If AA and CS attempt to decrypt partial ciphertext, they will have to solve the discrete logarithm problem.

\subsubsection{Verifiability}
Our vFAC encrypts a random key R using access policy, while the real message $M$ is hidden by symmetric algorithm $SE$ and symmetric key ${K_{SE}}$. Therefore, the verification of $V{K_M}$ ensures the correctness of the random key $R$, which is to ensure the correctness of the ciphertext decrypted by the CS.

\subsection{Performance Analyses}
%\subsubsection{Feature Comparison}\
%\emph{1) Features Comparison }
\subsubsection{Comparison of Features}
%In TABLE~\ref{tabl:comparison}, we compare the features between several schemes \cite{RW15,MZL17,YMCZZ17, LLL17,NMSM17}.
Table ~\ref{tabl:comparison} shows the comparison on features among the selected schemes. YMCZZ in \cite{YMCZZ17} is accountable, but its method of solving the key escrow problem will cause the waste of resources, which is difficult to implement under actual conditions. Besides, the schemes in \cite{YMCZZ17}, \cite{LLL17}, \cite{NMSM17} may have the problem of system construction if too many attributes are added to attribute authorities due to lack in the feature of large universe. Because our proposed vFAC provides all the features listed in table ~\ref{tabl:comparison}, it is more comprehensive than other schemes.

\subsubsection{Comparison of Computation Overhead}
We make a comparison on the phases of offline/online encryption and user decryption between vFAC and the selected schemes in TABLE~\ref{tab2:computational}. Let $l$ denote the number of rows of the access matrix, $\left| I \right|\left( {\left| I \right| \le l} \right)$ denote the number of rows used for decryption in the access matrix, $P$ denote bilinear pair operation, and $E$ denote exponential operation.

\begin{table}[!h]
\scriptsize
\caption{Comparison of Computation Overhead}\label{tab2:computational}
\begin{center}
\begin{tabular}{c|  c c |c   }
%\toprule 画顶部的线
  \hline \hline

\multirow{2}{*}{\textbf{Schemes}} &
 \multicolumn{2}{c|}{\textbf{Encryption}} &
 \multirow{2}{*}{\textbf{Decryption.user}} \\

\cline{2-3} & \raisebox{-0.05cm}[0pt]{Offline.Enc}& \raisebox{-0.05cm}[0pt]{Online.Enc}\\

  %\multirow{2}{*}\raisebox{-0.06cm}[0pt]{\textbf{Schemes}} & \raisebox{-0.06cm}[0pt]{ \textbf{Offline.Enc }} & \raisebox{-0.06cm}[0pt]{\textbf{Online.Enc}} & \multirow{2}{*}\raisebox{-0.06cm}[0pt]{\textbf{Encryption.user}}  \\

 \hline \hline
  \raisebox{-0.05cm}[0pt]{RW\cite{RW15}} & \multicolumn{2}{c|}{ \raisebox{-0.05cm}[0pt]{$(4l + 1)E$}} &{$3\left| I \right|P + 2\left| I \right|E$} \\

  \hline
  \raisebox{-0.05cm}[0pt]{MZL\cite{MZL17}} & \multicolumn{2}{c|}{\raisebox{-0.05cm}[0pt]{$(4l + 1)E$}} & \raisebox{-0.05cm}[0pt]{$E$} \\

 \hline
  \raisebox{-0.05cm}[0pt]{YMCZZ\cite{YMCZZ17}} & \multicolumn{2}{c|}{ \raisebox{-0.05cm}[0pt]{$(4l + 1)E$}} & \raisebox{-0.05cm}[0pt]{$ 3\left| I \right|P + 2\left| I \right|E  $}  \\
 \hline
  \raisebox{-0.05cm}[0pt]{LLL\cite{LLL17}} & \raisebox{-0.05cm}[0pt]{$3lE$} & \raisebox{-0.05cm}[0pt]{$2E$} & \raisebox{-0.05cm}[0pt]{$E$} \\
 \hline
  \raisebox{-0.05cm}[0pt]{NMSM\cite{NMSM17}} & \multicolumn{2}{c|}{ \raisebox{-0.05cm}[0pt]{$(l + 2)E$}} & \raisebox{-0.05cm}[0pt]{$(\left| I \right| + 1)P + (\left| I \right| + 1)E$} \\

 \hline
  \raisebox{-0.05cm}[0pt]{vFAC} & \raisebox{-0.05cm}[0pt]{$4lE$} & \raisebox{-0.05cm}[0pt]{$2E$} & \raisebox{-0.05cm}[0pt]{$E$} \\

%\toprule
  \hline \hline

\end{tabular}
\end{center}
\end{table}
%\vspace{-0.8em}
In TABLE~\ref{tab2:computational}, we found that the schemes in \cite{RW15,MZL17,YMCZZ17,NMSM17} have no offline encryption mechanism, which cause the number of operations in encryption phase linearly increasing with $l$. In \cite{RW15,YMCZZ17,NMSM17}, the user directly decrypts the original ciphertext, so the exponential operations and the number of bilinear pair operations also linearly  increase with $\left| I \right|$ in the decryption phase, leading to high computational complexity. In the decryption process of vFAC, only one exponent operation is involved. Although, in \cite{LLL17}, the user also only needs one decryption operation, it is achieved by outsourcing decryption and cannot solve the key escrow problem. For all the above, vFAC has high computational efficiency on the user side.

\subsubsection{Comparison of Storage Cost}
%\emph{3) Comparison of Storage Cost}
%\subsubsection{ Comparison of Storage Cost}\

\begin{table*}[!ht]
\scriptsize
\caption{Comparison of Storage Cost}\label{tab3:storage}
%\caption{Comparison of Storage Cost}
\begin{center}
\begin{tabular}{c  c c c c   }
%\toprule
  \hline \hline

  \raisebox{-0.06cm}[0pt]{\textbf{Schemes}} & \raisebox{-0.06cm}[0pt]{ \textbf{Secret key of AA} } & \raisebox{-0.06cm}[0pt]{\textbf{Public key of AA} } & \raisebox{-0.06cm}[0pt]{\textbf{Private key of user} } & \raisebox{-0.06cm}[0pt]{\textbf{Ciphertext size }} \\
%\toprule
 \hline \hline
  \raisebox{-0.05cm}[0pt]{RW\cite{RW15}} & \raisebox{-0.05cm}[0pt]{$2\left| {{Z_{\rm{p}}}} \right|$} & \raisebox{-0.05cm}[0pt]{$2\left| G \right|$} & \raisebox{-0.05cm}[0pt]{$2\left| G \right|\left| S \right|$} & \raisebox{-0.05cm}[0pt]{$(4l + 1)\left| G \right|$} \\

  %\hline
  \raisebox{-0.05cm}[0pt]{MZL\cite{MZL17}} & \raisebox{-0.05cm}[0pt]{$2\left| {{Z_{\rm{p}}}} \right|$} & \raisebox{-0.05cm}[0pt]{$2\left| G \right|$} & \raisebox{-0.05cm}[0pt]{$\left| {{Z_{\rm{p}}}} \right|$} & \raisebox{-0.05cm}[0pt]{$(4l + 1)\left| G \right|$}  \\

 % \hline
  \raisebox{-0.05cm}[0pt]{YMCZZ\cite{YMCZZ17}} & \raisebox{-0.05cm}[0pt]{$ 2\left| {{Z_{\rm{p}}}} \right| $} & \raisebox{-0.05cm}[0pt]{$2\left| G \right|$} & \raisebox{-0.05cm}[0pt]{$ 2\left| G \right|\left| S \right| $} & \raisebox{-0.05cm}[0pt]{$(4l + 1)\left| G \right|$}  \\

  \raisebox{-0.05cm}[0pt]{LLL\cite{LLL17}} & \raisebox{-0.05cm}[0pt]{$1 + \left| U \right|\left| {{Z_{\rm{p}}}} \right|$} & \raisebox{-0.05cm}[0pt]{$1 + \left| U \right|\left| G \right|$} & \raisebox{-0.05cm}[0pt]{$2 + \left| S \right|\left| G \right| + \left| {{Z_{\rm{p}}}} \right| $} & \raisebox{-0.05cm}[0pt]{$(3l + 2)\left| G \right| + 2l\left| {{Z_{\rm{p}}}} \right| + \left| c \right| + \left| {VK} \right|$}  \\

  \raisebox{-0.05cm}[0pt]{NMSM\cite{NMSM17}} & \raisebox{-0.05cm}[0pt]{$(5\left| U \right| + 1)\left| {{Z_{\rm{p}}}} \right|$} & \raisebox{-0.05cm}[0pt]{$ (3\left| U \right| + 1)\left| G \right| $} & \raisebox{-0.05cm}[0pt]{$ (2\left| S \right| + 1)\left| G \right| $} & \raisebox{-0.05cm}[0pt]{$ (l + 2)\left| G \right| $}  \\

  \raisebox{-0.05cm}[0pt]{vFAC} & \raisebox{-0.05cm}[0pt]{$3\left| {{Z_{\rm{p}}}} \right|$} & \raisebox{-0.05cm}[0pt]{$3\left| G \right|$} & \raisebox{-0.05cm}[0pt]{$2\left| {{Z_{\rm{p}}}} \right|$} & \raisebox{-0.05cm}[0pt]{$(4l + 2)\left| G \right| + 2l\left| {{Z_{\rm{p}}}} \right| + \left| c \right| + \left| {VK} \right|$}  \\

%\toprule
  \hline \hline

\end{tabular}
\end{center}
\end{table*}
Denote $\left| {{Z_p}} \right|$ and $\left| G \right|$ as the length of element in ${Z_p}$ and $G$, $\left| U \right|$ as the number of attributes managed by AA, $\left| S \right|$ as the number of user's attributes,  $\left| c \right|$ as the length of the ciphertext after symmetric encryption, and $\left| {VK} \right|$ as the length of the verification key. The Table~\ref{tab3:storage} shows the comparison result of storage cost. The length of ciphertext is linearly related to $\left| I \right|$ because the ciphertext corresponds to the access policy. The storage capacity of the CS is actually stronger than that of users. Therefore, the storage cost of ciphertext on the CS can be omitted. Here, we mainly focus on the user's storage cost.

In \cite{LLL17,NMSM17}, the length of the public/private key of each AA is linearly related to the number of its attributes. Therefore, the length of public/private key of AA is linearly related to $\left| U \right|$. The users private key in \cite{RW15,YMCZZ17,LLL17,NMSM17} is directly generated by AA based on the user's attributes, so the length of private key is linearly related to $\left| S \right|$. When the number of attributes increases, the user's storage cost increases too. In vFAC, the length of AA's public/private key is a fixed value because it is independent of the number of attributes. Although the whole storage cost in \cite{YMCZZ17} is lower than vFAC, it does not meet the property of large universe. In conclusion, vFAC is more suited for data access control because of its comprehensive features.

\section{Conclusion}
In order to solve the fine-grained data access control problem in cloud storage, this paper proposes a fine-grained access control scheme for cloud storage based on multi-authority CP-ABE. The proposed vFAC not only realizes online/offline encryption mechanism, but also satisfy the feature of hidden policy. Furthermore, vFAC allows the user to verify the decrypted ciphertext to ensure that the CS decrypts ciphers correctly. The static security of  vFAC is also proved under the random oracle model. In particular, the analyses of features, computation overhead, and storage cost with the other existing schemes show that the vFAC has a more comprehensive advantage for cloud storage.

\section*{Acknowledgements}
This work is supported by the Key Program of NSFC-Tongyong Union Foundation under Grant U1636209, the 111 Project (B08038) and Collaborative Innovation Center of Information Sensing and Understanding at Xidian University.

\bibliographystyle{IEEEtran}
\bibliography{ms}
\end{document}